\documentclass[useAMS,usenatbib]{mn2e}
\usepackage{graphicx,fleqn}
\DeclareGraphicsExtensions{.eps,.ps,.eps.gz,.ps.gz,.eps.Z}
\title[STJ Spectrograph]{A Concept for an STJ-based Spectrograph}

\author[]{Mark Cropper$^{1}$, M. Barlow$^{2}$, M. A. C. Perryman$^{3}$,
Keith Horne$^{4}$, R. Bingham$^{2}$,\and M. Page$^{1}$ , P. Guttridge$^{1}$, 
A. Smith$^{1}$, A. Peacock$^{3}$, D. Walker $^{2}$ and P. Charles$^{5}$\\
$^{1}$Mullard Space Science Lab, University College London,
Holmbury St. Mary, Dorking, Surrey, RH5 6NT, UK\\
$^{2}$ Department of Physics and Astronomy, University College London, Gower
Street, London WC1E 6BT\\
$^{3}$ Research and Scientific Support Department of ESA, ESTEC, Posbus 299, 
Keplerlaan 1, Noordwijk NL-2200 AG, The Netherlands\\
$^{4}$ School of Physics and Astronomy, University of St. Andrews, North Haugh,
St. Andrews, Fife, Scotland \\
$^{5}$ Department of Physics and Astronomy, University of Southampton,
Hampshire SO17 1BJ}

\date{Received: }

\begin{document}

\newcommand{\dg} {^{\circ}}
\outer\def\gtae {$\buildrel {\lower3pt\hbox{$>$}} \over
{\lower2pt\hbox{$\sim$}} $}
\outer\def\ltae {$\buildrel {\lower3pt\hbox{$<$}} \over
{\lower2pt\hbox{$\sim$}} $}
\newcommand{\ergscm} {ergs s$^{-1}$ cm$^{-2}$}
\newcommand{\ergss} {ergs s$^{-1}$}
\newcommand{\ergsd} {ergs s$^{-1}$ $d^{2}_{100}$}
\newcommand{\pcmsq} {cm$^{-2}$}
\newcommand{\ros} {{\it ROSAT}}
\newcommand{\xmm} {{\it XMM-Newton}}
\newcommand{\exo} {{\it EXOSAT}}
\newcommand{\sax} {{\it BeppoSAX}}
\newcommand{\chandra} {{\it Chandra}}
\def\rchi{{${\chi}_{\nu}^{2}$}}
\newcommand{\Msun} {$M_{\odot}$}
\newcommand{\Mwd} {$M_{wd}$}
\def\Mdot{\hbox{$\dot M$}}
\def\mdot{\hbox{$\dot m$}}
\def\mincir{\raise -2.truept\hbox{\rlap{\hbox{$\sim$}}\raise5.truept
\hbox{$<$}\ }}
\def\magcir{\raise -4.truept\hbox{\rlap{\hbox{$\sim$}}\raise5.truept
\hbox{$>$}\ }}

\maketitle

\begin{abstract}
We describe a multi-order spectrograph concept 
suitable for 8m-class telescopes,
using the intrinsic spectral resolution of Superconducting Tunneling Junction
detectors to sort the spectral orders. The spectrograph works at low orders,
1--5 or 1--6, and provides spectral coverage with a resolving power of
R$\simeq8000$ from the atmospheric cutoff at 320 nm to the long wavelength end
of the infrared H or K band at 1800 nm or 2400 nm. We calculate that the
spectrograph would provide substantial throughput and wavelength coverage, 
together with high time
resolution and sufficient dynamic range. The concept uses currently available
technology, or technologies with short development horizons, restricting the
spatial sampling to two linear arrays; however an
upgrade path to provide more spatial sampling is identified. All of the other
challenging aspects of the concept -- the cryogenics, thermal baffling and
magnetic field biasing -- are identified as being feasible.
\end{abstract}

\section{Introduction}
\label{int}

The scientific case for a medium resolution spectrograph with high throughput,
and covering both optical and near infrared, is substantial. A brief resume
would include time-resolved studies of planetary transits, where spectra taken
at ingress and egress provide information on the planetary atmospheres;
$\gamma$-ray bursters, where the spectral evolution during the fading afterglow
provides an indication of the interaction of the fireball with its
surroundings, and therefore information on the progenitor; studies of early
structure and pre-galactic clouds through spectra of absorption line systems in
distant quasars; source identifications and redshifts of star-forming galaxies
at high redshift detected in submillimetre surveys; supernovae and the
interaction between their ejecta and previous episodes of mass loss, leading to
estimates of chemical abundances, density distributions and ejecta masses;
novae and accreting binary stars, where indirect techniques such as echo and
eclipse mapping and Doppler tomography provide detailed information on
accretion flows onto neutron stars, black holes and white dwarfs at
micro-arcsecond scales; echo mapping of accretion in active galaxies and in
microlensing studies, where detailed information can be obtained on both lensed
and lensing object from the time-resolved spectra. All of these studies require
only a small field of view and a spectral resolving power $>5000$; high time
resolution is also important. The coverage of both optical and infrared bands
is essential, not only in measuring the spectral energy distribution as it
evolves, but also in the relationship between spectral lines at widely
different wavelengths, and in the accessibility of the redshifted UV and
optical lines in objects at high redshift.

Consequently, one of the instruments identified for a second generation ESO VLT
instrument suite was a spectrograph of medium resolving power 
($R\sim 10\,000$),
with a small field of view (arcsec) and spectral coverage over almost a decade
in wavelength from the atmospheric cutoff at 320 nm to the long wavelength end
of the K band at 2400 nm. The capability to take relatively short exposures was
also cited as important, and the instrument was required to deliver a factor
$>2.5$ improvement in throughput over current instrumentation. This study was
carried out in preparation for a response to ESO, but not taken further owing
to changes in the manufacturing capability for the detectors.

An increase in throughput by a significant factor over existing front-rank
instrumentation is challenging. One improvement is to replace the
cross-dispersing grating in a classical echelle with prisms, as in UCLES at the
AAT (Walker \& Diego 1985) and bHROS on Gemini (Diego et al 1997). This should
increase the throughput of the cross-dispersing optics by a factor of perhaps
1.5, and also reduce the polarisation effects
between dispersing elements suffered in cross-dispersed
spectrographs. Alternatively, a non-echelle spectrograph could be feasible,
with dichroics splitting the optical into two beams and three beams in the
infrared, where the detector format is more problematic.  This beamsplitting
allows the optics and gratings to be optimised for each beam, but is
inefficient, and allows spectral leakage. Consequently, the overall gains are
likely to be modest, and such a spectrograph would be complex.

In order to achieve the increased performance required, a radical, but
optically simple design is needed. 
We describe such a concept here, 
a multi-order spectrograph that takes advantage of the intrinsic colour
resolution of Superconducting Tunnelling Junction (STJ) detectors.
STJs are the first detectors to be able to intrinsically distinguish
colour in the UV-IR band (Perryman et al 1993), and, in their Tantalum
form, have sufficient wavelength resolution to separate the spectral orders, 
eliminating the need for order-sorting
optics. We calculate that with 5000-pixel linear arrays of STJ
detectors, most of the 320--2400 nm range can be covered at a resolving
power $R\sim10\,000$ with orders 1--5 or 1--6, using a single detector
system. The simplicity of the design enables a higher throughput to be
reached.  STJs have a sensitivity exceeding 70\% (uncoated) through
most of the optical band. This drops to a few percent into the
infrared, but the coverage of the entire band compensates for the
lower quantum efficiency of the detectors.

Because STJ detectors are photon-counting, high time resolution
information is preserved. The exposure time can be constructed
non-destructively {\it postfacto}, at the data analysis stage. This is
particularly important for variable objects, such as $\gamma$-ray
bursters, where the intensity changes are not predictable beforehand.
It also eliminates the overheads of readout time, which can be very
large in the case of non-frame-transfer CCDs for short exposure times,
especially for echelle formats: this in itself can lead to factors up
to 2 increase in throughput, for short exposures.

A further quality of STJs is that they have no readout noise, and
insignificant dark noise. Despite the low readout noise of the best
current-generation CCDs, this still causes a significant reduction in
S/N ratio in CCD echelle spectrographs. The gain is particularly
marked in the infrared, where current infrared detector technology
cannot hope to match the STJ performance in this respect.

Larger format STJ arrays could be envisaged for the future, but to
keep within currently feasible technology we restrict this instrument
concept to two linear arrays (the second for sky
subtraction). This means that the spectrograph will be optimised for
sources of small angular extent, the effective field of view being set by
the spatial extent of each pixel. It may, however, be possible to
provide some spatial resolution as an upgrade path relatively simply.

Initial studies for an STJ-based spectrograph working in the UV were presented
in the HSTJ proposal for the Hubble Space Telescope (Griffiths et al 1997). A
more detailed exposition of this concept, including a full science case, can be
found in Cropper et al (2002).

\section{Instrument design}

\subsection{An overview of STJ detectors}

STJ detectors were first developed for the X-ray band, but their
potential for use in the UV/optical/IR band was realised by Perryman
et al (1993). They are the first detectors to provide intrinsic colour
discrimination at optical wavelengths. This is possible because the
energy to break the Cooper-pairs in the superconductor is small
compared to the energy of an optical photon. The details of the
operation of STJs can be found in Perryman et al (1993), but the
essence is that more or fewer electrons are generated depending on
whether the incident photon is blue or red. It then remains to measure
this charge, to determine its wavelength.

\begin{figure}
\begin{center}
\includegraphics[scale=0.4]{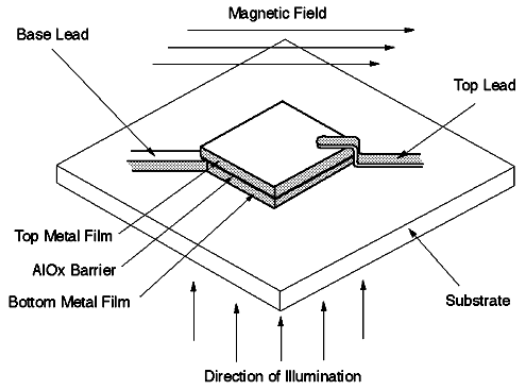}\\
\vspace*{2mm}
\includegraphics[scale=0.33]{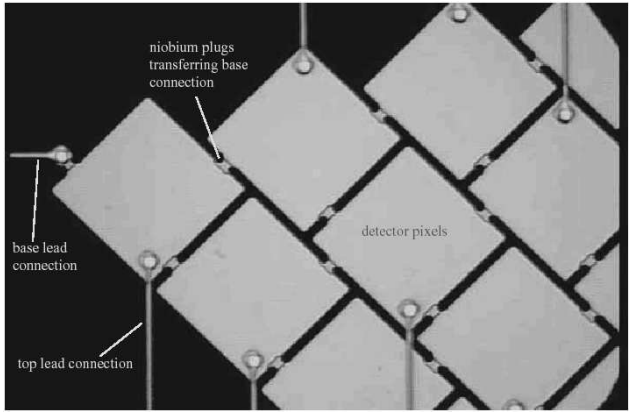}
\end{center}
\caption{(upper) A schematic of an STJ pixel. (lower) A detail image of the $6\times6$
Tantalum STJ array currently in use in S-Cam2. }
\label{fig:STJ_pixel}
\end{figure}

An STJ consists of two metal films, separated by an Aluminium Oxide
barrier, across which the electrons tunnel. This `sandwich'
constitutes a pixel. It is supported on a substrate, such as sapphire,
though which it is illuminated. Each pixel requires its own electrical
connection: generally the bottom metal film is connected in common
with the other pixels, while the top metal film has a unique
connection. The device requires a magnetic field bias to suppress the
Josephson current. A schematic is given in Figure~\ref{fig:STJ_pixel}.

Current generation STJs use Tantalum metal films. Pixel sizes range from
$20-50\mu$m. The STJ arrays in the S-Cam2 instrument in use at the 4.2m William
Herschel Telescope at La Palma have a format of $6\times6$ Tantalum pixels in a
staggered rectangular array (Perryman et al 2001 and references therein). Each
pixel has its own independent preamplifier and analog electronics chain. This
array has now been in use for more than two years, and superseded a first
generation array of the same format. S-Cam2 provides a spectral resolving power
of $R\simeq8$ over the 300--650 nm band. The timing accuracy for the photon
events is $5\mu$s, and the maximum count-rate is 5000 counts/pixel/sec. A new
$10\times12$ pixel array for the camera is under test, and developments using
other materials to provide higher intrinsic spectral resolution are under
way. Scientific results from S-Cam include the determination of QSO redshifts
(de Bruijne et al 2002), studies of the accretion regions and streams in
eclipsing Cataclysmic Variables (Perryman et al 2001, Bridge et al 2002a,b),
and observations of the Crab pulsar (Perryman et al 1999).

\subsection{Detector constraints}

\subsubsection{Array size and Pixel Layout}

To reach the wavelength coverage required, the spectrograph will need the
maximum array length in the dispersion direction. The limits here are set by
the wafer size, which we assume nominally to be 5 cm, together with the minimum
practical pixel size, $20\mu$m, giving a maximum length of 2500 pixels. This is
insufficient (see section~\ref{sec:concepts}), so two such arrays must be
butted to give 5000 pixels in the dispersion direction.

While more than two arrays could be used, thus providing increased wavelength
coverage, other aspects (optics, thermal issues) rapidly become more
problematic. Because in this concept we have maintained the principle that at
most only modest technological developments should be required for the
realisation of the concept, we reserve such enhancements for the future and
limit the array lengths to 5000 pixels.

Yields for current STJ devices are sufficiently high to give confidence that
each 2500 pixel strip can be fabricated successfully, so that our concept is
feasible. The join region can be made small. Moreover, one of the major
difficulties limiting the size of {\it square} STJ arrays is the access for
electrical connections to each pixel: in {\it linear} arrays this is much
easier. Cross-talk betwen pixels is also lower for the linear arrays since each
pixel has half the number of adjacent pixels.

The spectrograph concept assumes a `small' field of view (arcsec). The most
conservative approach would be to use two strips of detector pixels, one for
star and one for sky subtraction, each only one pixel wide. This results in a
total of $10\,000$ pixels, each with easy 
access for electrical connections. This
number of pixels is acceptable in terms of electrical connections and signal
processing chains (see section~\ref{sec:concepts_electronics}). There is some freedom
in the width (in the spatial direction) of the pixels. Such rectangular pixels
may also slightly increase the intrinsic STJ wavelength resolution without a
concommitant increase in array length.

Some spatial information along the slit could be achieved using DROIDs. These
could be introduced in our concept at a later upgrade path, as their technology
matures. DROIDs (Verhoeve et al 2000) consist of two STJs placed at the end of
a Tantalum detection strip; the summed charge is related to the energy of the
photon, while the location at which the photon was detected can be recovered
from the difference in the two charges collected by the STJs. Verhoeve et al
(2000) find that $20\times200\mu$m devices provide an effective 11 pixels, for
just two electrical connections. The issues here are whether the
DROIDs can be laid down efficiently in such large linear arrays, their spectral
resolution is sufficient for the order separation (for the spectral formats
below this appears to be the case), and whether it is more desirable
scientifically to have the sky subtraction array further displaced from the
source array than is possible over the effective 11 pixels provided by the
DROID.

\subsubsection{Intrinsic Resolution}

The key to the spectrograph concept is that the intrinsic wavelength
resolution of the STJs is used to perform the order sorting.

Tantalum STJs with $20\mu$m pixels have a theoretical wavelength resolving
power $\lambda/\Delta\lambda$ of 20 at 320 nm (Peacock et al 1998), decreasing
with wavelength to 8.1 at 2400 nm. The resolving power depends on the pixel
size: $50\mu$m pixels have slightly better resolving power of 22 at 320 nm. In
the S-Cam2 array, a practical resolving power of 12 is reached at 320 nm 
(Rando et al 2000b), limited by infrared background and electronics
noise. (The spectral resolving power of DROIDs is $\sim10-50$\% poorer
depending on pixel size, worse for larger pixels.) 


Other materials such as Hafnium and Molybdenum provide superior
spectral resolving power (eg Verhoeve et al 2000), with Hafnium
exceeding 100 at 320 nm. However although some junctions have been
fabricated from such materials, the technology is not yet sufficiently
mature to use in this concept, and the spectral resolution of Tantalum
is in any case sufficient.

\subsubsection{Count Rate and Time Resolution}

STJs are photon-counting, and intrinsically fast devices. In practise,
their time resolution is set by the pulse-counting electronics and the
resolution of the time-tagging electronics. In S-Cam, this accuracy is
in the regime of tens of microseconds.

The S-Cam2 devices count up to in excess of 5000 counts/sec/pixel. The detector
resolution degrades slightly with countrate ($\sim15$\% at 5000 c/s/pix),
mostly due to pileup in the currently realised pulse height analysing
electronics (Rando et al 2000b). Faster countrates would be possible, with
revised electronics; however for this concept we take the approach that the
current countrates of 5000 should provide an adequate working count-rate limit
for sufficient dynamic range.

\subsubsection{Quantum Efficiency}

The quantum-efficiency of STJs exceeds 70\% in the U--R bands, and
drops towards longer wavelengths, reaching 20\% at 1000 nm and 5\% at
2400 nm (see (Peacock et al 1998, Verhoeve et al 2000.).
This is the result mainly of reflection off the Tantalum film. Other (less
well developed) STJ materials, Hafnium and Molybdenum, have reflective
qualities similar to Tantalum in the infrared. From the point of
view of infrared sensitivity, there are therefore no particular gains to be
made from using them.


The infrared reflectance can potentially be improved by the deposition
of an appropriate coating on the substrate, before the pixel structure
is laid down. Currently, little work has been done to investigate this
possibility, but this would be obviously desirable for this
spectrograph. 

The steep drop in quantum efficiency beyond $5\mu$m in the infrared
aids in ensuring the greater thermal background flux at longer wavelengths
does not saturate the detector countrate.

\subsection{Spectrograph Concepts}
\label{sec:concepts}

\subsubsection{Resolution requirements for order sorting}

In order to sort order 2 from 1, a detector needs to have a resolving power of
2. Generalising, the highest order a detector can sort is set by the resolving
power of the detector at that wavelength.

The increase in order number with decreasing wavelength implies that the
resolving power needs to be highest at the extreme blue end of the operating
range. The resolution of the STJs increases with decreasing wavelength. This is
in the right sense to match the order sorting requirement. In order to
minimise the possibility of allocating a photon to the incorrect order, we need
some resolution margin $f_r\sim1.5-2$. If we use the current S-Cam2 resolving
power of $\sim10-12$ at 320 nm, then for $f_r\sim2$ the maximum usable order is
$5-6$. For these low order numbers the resolving power required to sort lower
orders drops more quickly than the intrinsic resolving power of the STJ, so all
lower orders can also be sorted (and more easily).


\subsubsection{Basic constraints from the order sorting}
\label{sec:order_sorting}

The basic constraints on multi-order spectrographs are simply derived from
first principles. We start from the grating equation:
\begin{equation}
m\lambda = d(\sin{i}+\sin{\theta})
\label{eq:grating}
\end{equation}
where $m$ is the order number, $d$ is the grating spacing ({\it i.e.}
$1/d$ is the number of grooves per unit length), $i$ is the angle of
the incident ray, and $\theta$ is the angle of the diffracted ray.
Generally $i$ and $d$ are constant. This indicates that at any
position on the detector (implying constant $\theta$) for all
orders, $m\lambda=$ constant. So for example, if a particular pixel
in order 1 corresponds to 1000 nm, then light from 500 nm in order 2
will also fall on the pixel, 333 nm in order 3, and so on.

If in our case we allow the maximum order to be 5, with a central
wavelength of 350 nm, then order 1 will be at 1750 nm. Alternatively,
if we wanted a central wavelength of 2100 nm, with coverage down to
350 nm, then the highest order will be 6.

\subsubsection{Dispersion}

The actual wavelength range covered in each order is determined by
differentiating the grating equation with respect to
wavelength:
\begin{equation}
\frac{\delta\theta}{\delta\lambda} = \frac{m}{d\cos{\theta}}.
\label{eq:grating_dif}
\end{equation}
Here $\delta\theta$ may, for example, correspond to one pixel on the
detector. Thus $\delta\lambda/\delta\theta$ corresponds to the
dispersion. We can see that at a particular position on
the detector (again, a constant $\theta$), the dispersion,
$\delta\lambda/\delta\theta$, is proportional to $1/m$. 

If we calculate the resolving power $R=\lambda/\delta\lambda$, then at
a particular position on the detector, with $\lambda=\lambda_1/m$
($\lambda_1$ is the wavelength in first order), and $\delta\lambda =
C/m$ ($C$ a constant), then $R=\lambda_1/C$, {\it i.e.} constant,
independent of order number. Thus although the numerator in
$R=\lambda/\delta\lambda$ is halved, so has the step in wavelength
corresponding to, say, one pixel. The consequence of this is that all
orders have the same resolving power at a particular pixel on the
detector.

\subsubsection{Spectral length}
\label{sec:spectral_length}

The next step is to calculate the length of the spectrum. {\it Within}
each order, Equation~\ref{eq:grating_dif} indicates that the
dispersion is approximately constant if $\cos\theta$ does not change
too much (which is generally true). Thus, within each order, each pixel
corresponds to a fixed wavelength interval. Now if we need a resolving
power $R$ corresponding to $2$ pixels (for Nyquist sampling of the
slit), then the dispersion
\begin{equation}
\delta\lambda = \lambda_c/2R
\end{equation}
where $\lambda_c$ is the centre wavelength of the first order. For example, if
we require a resolving power of $10\,000$ at 2000 nm, this corresponds to 0.2
nm/resolution element, or 0.1 nm/pix (Nyquist). 

The wavelength coverage in each order is then simply set by the number
of pixels in the detector. For the above example, a 5000 pixel
detector provides coverage in first order of $5000\times0.1=500$ nm
centred on 2000 nm, {\it i.e.} $1750-2250$ nm. In second order it
provides $5000\times0.05=250$ nm centred on 1000 nm, so $875-1125$ nm,
and so on. It is evident in this example that there is a gap between
1125 and 1750 nm not covered by the detector. Gaps can be fixed either
by lowering the required resolution, or increasing the detector
length. Neither of these may be feasible, in which case the location
of the gaps has to be optimised.

\subsubsection{Blaze efficiency}

The blaze curve of a grating is approximately unchanged with order if
plotted as a function of $m\lambda$, instead of $\lambda$. The
efficiency falls off from the blaze, typically reaching half the peak
efficiency at $2/3(\lambda_b/m)$ and $3/2(\lambda_b/m$), where
$\lambda_b/m$ is the blaze wavelength for order $m$. 
If we adopt the criterion that the length of
each order is defined by the wavelengths at which the efficiency drops
to below that of prior or following orders, then the spectrum appears
to shorten gradually as the orders (and thus order overlap) increase.
This is familiar from the raw images taken by cross-dispersed echelle
spectrographs. It is important to note, however, that some photons
will arrive in a particular order beyond this wavelength, and will be
there for collection if there are active detector pixels to detect
them.

\subsubsection{Possible Configurations}

We now apply these considerations with the following inputs: the maximum order
is $\sim5-6$, set by the STJ resolving power with factor $f_r$; the minimum
wavelength should be the atmospheric cutoff at 320 nm; the resolving power
should be $\sim10\,000$ and the detector length should be limited to 5000
pixels, set by realistic expectations of array size.

\begin{figure}
\begin{center}
\includegraphics[scale=0.5,angle=-90]{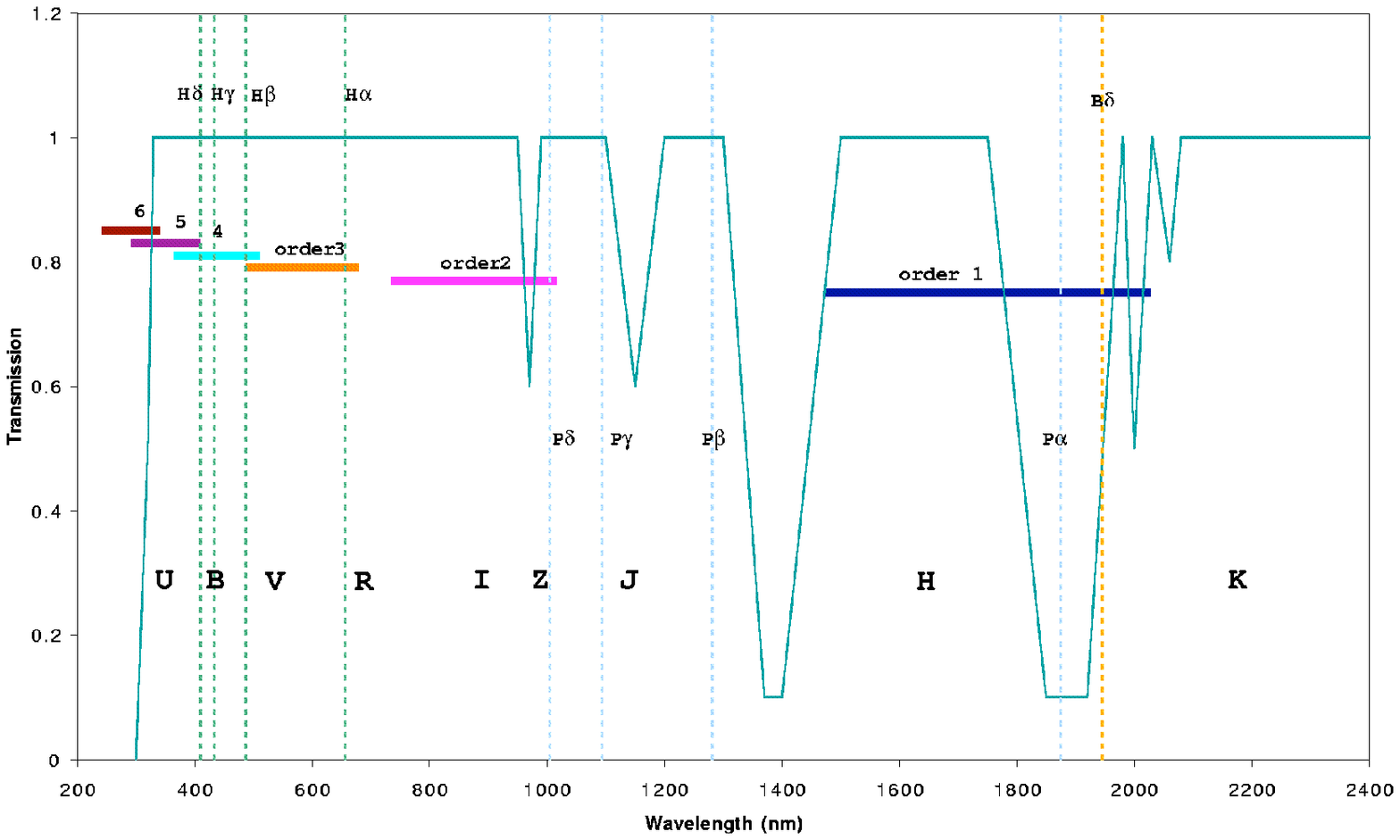}
\includegraphics[scale=0.5,angle=-90]{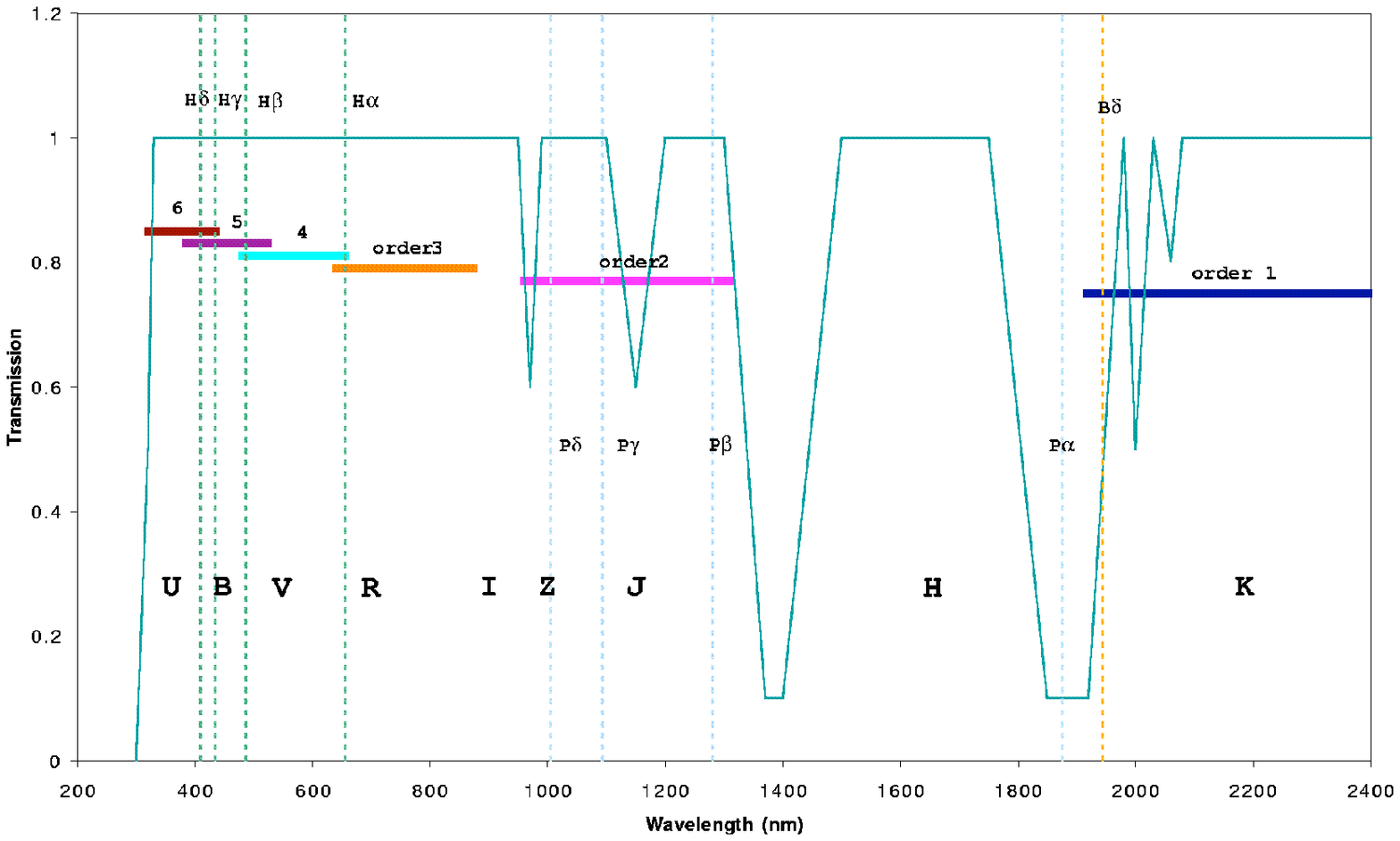}
\end{center}
\caption{The order coverage in the baseline H band configuration (upper)
and alternative K band configuration (lower), together with a schematic
depiction of the atmospheric transparency. Also shown are the approximate
wavelengths of the U--K photometric bands and some of the Hydrogen Balmer,
Paschen and Brackett lines.}
\label{fig:echelle_pattern}
\end{figure}

We have given an example in section~\ref{sec:spectral_length} above with a
resolving power of of $10\,000$. This 
has gaps between orders for orders up to 4. A
reduction in resolving power to 8000 closes the gap between order 3 and 4, and
leaves only those between 1 and 2 and between 2 and 3 (only 50 nm around 700 nm
in this latter case). Further adjustments can be made in the resolving power,
and in the centre wavelength of the first order, but assuming
$\lambda_c(1)=1750$ nm and $R(\lambda_c)=8000$ then the spectral coverage in
Figure~\ref{fig:echelle_pattern}(upper) is obtained.  This has the advantage of
covering the U--Z bands and the H band in the infrared.

For a highest order 6, the centre wavelength of $m=1$ is 2100 nm, in the K
band, and the coverage is now U--R, Z, J and K bands, missing H and Z. This is
shown in Figure~\ref{fig:echelle_pattern}(lower). The wavelengths covered in 
each band are shown in Table~\ref{tab:echelle_orders} .

\begin{table*}
\begin{center}
\begin{tabular}{rrrrrrrr}
Order & $\lambda_c$ & $\Delta\lambda_c$ & Resolving & $\lambda_r$ &
$\lambda_b$ & Overlap & Overlap \\ 
      & (nm)        & (nm)              & Power R$_c$ & (nm)        &
(nm)        & (nm)    & (pix)   \\ \hline
1 &     1750 &  0.1094  & 8000 & 2023 &   1477 & -465 &   -4250  \\
2 &     875  &  0.0547  & 8000 & 1012 &    738 &  -64 &   -1167  \\
3 &     583  &  0.0365  & 8000 &  674 &    492 &   14 &     375  \\ 
4 &     438  &  0.0273  & 8000 &  506 &    369 &   36 &    1300  \\
5 &     350  &  0.0219  & 8000 &  405 &    295 &   42 &    1917  \\
\hline\\[5mm]
Order & $\lambda_c$ & $\Delta\lambda_c$ & Resolving & $\lambda_r$ &
$\lambda_b$ & Overlap & Overlap \\ 
      & (nm)        & (nm)              & Power R$_c$ & (nm)        &
(nm)        & (nm)    & (pix)   \\ \hline
1 &     2270 &  0.1419  & 8000 & 2625 &   1915 & -603 &   -4250  \\
2 &     1135 &  0.0709  & 8000 & 1312 &    958 &  -83 &   -1167  \\
3 &     757  &  0.0473  & 8000 &  875 &    638 &   18 &     375  \\
4 &     568  &  0.0355  & 8000 &  656 &    479 &   46 &    1300  \\
5 &     454  &  0.0284  & 8000 &  525 &    383 &   54 &    1917  \\
6 &     378  &  0.0236  & 8000 &  437 &    319 &   56 &    2357  \\
7 &     324  &  0.0203  & 8000 &  375 &    274 &   54 &    2688  \\
\hline

\end{tabular}
\end{center}
\caption{Details of the order coverage provided with a 5000 pixel array,
as shown graphically in Figure~\ref{fig:echelle_pattern}. $\lambda_r$ and
$\lambda_b$ are the blue and red extrema of each order, while the resolving
power is calculated for the central wavelength $\lambda_c$. $\Delta\lambda_c$
is the dispersion in nm/pix at $\lambda_c$. A negative overlap denotes a gap in
the spectral coverage.}
\label{tab:echelle_orders}
\end{table*}

Exploring further alternatives, if we place order 1 in the J band, the gap to
order 2 includes R and I bands, which is clearly unsatisfactory. On the other
hand, to cover J, H and K bands, the commensurability of their bandcentre
wavelengths is approximately in the ratio 4:3:2, so that order 2 should be
selected to be at $\sim2200$ nm. Unfortunately, this requires a large number of
orders, $\sim 20$ to reach $320$ nm, for which an STJ spectral resolving power
of $\sim 40$ would be required. This is out of the reach of Tantalum devices,
while insufficient experience has been gained with other materials such as
Molybdenum for their use in this concept. This arrangement of orders may be of
use in the future.

Of the earlier two possibilities, the first, with $\lambda_c(1)$ in the H band,
provides optimal and almost complete coverage through the optical band up to Z,
and complete coverage of H. In the second, where $\lambda_c(1)$ is in the K
band, an important consideration is the thermal infrared, making this more
challenging to realise as a practical design. The loss of some of the I band
may also be undesirable; on the other hand, the J band is gained. A
conservative approach would place $\lambda_c(1)$ in the H band, but further
consideration may well conclude that the larger infrared coverage provided by
$\lambda_c(1)$ in the K band is more desirable scientifically, and also
technically feasible.

\subsection{Optical design}
\label{sec:optical_design}

Our optical design concept is relatively simple: a multi-order spectrograph
without cross disperser, using an off-axis collimator and folded Schmidt
camera with an accessible focus. The main challenges here relate to the
infrared baffling and long wavelength supression, together with the
space required at the focus of the camera for the uniform magnetic
field system to bias the STJ array.

\begin{figure}
\begin{center}
\includegraphics[scale=0.33]{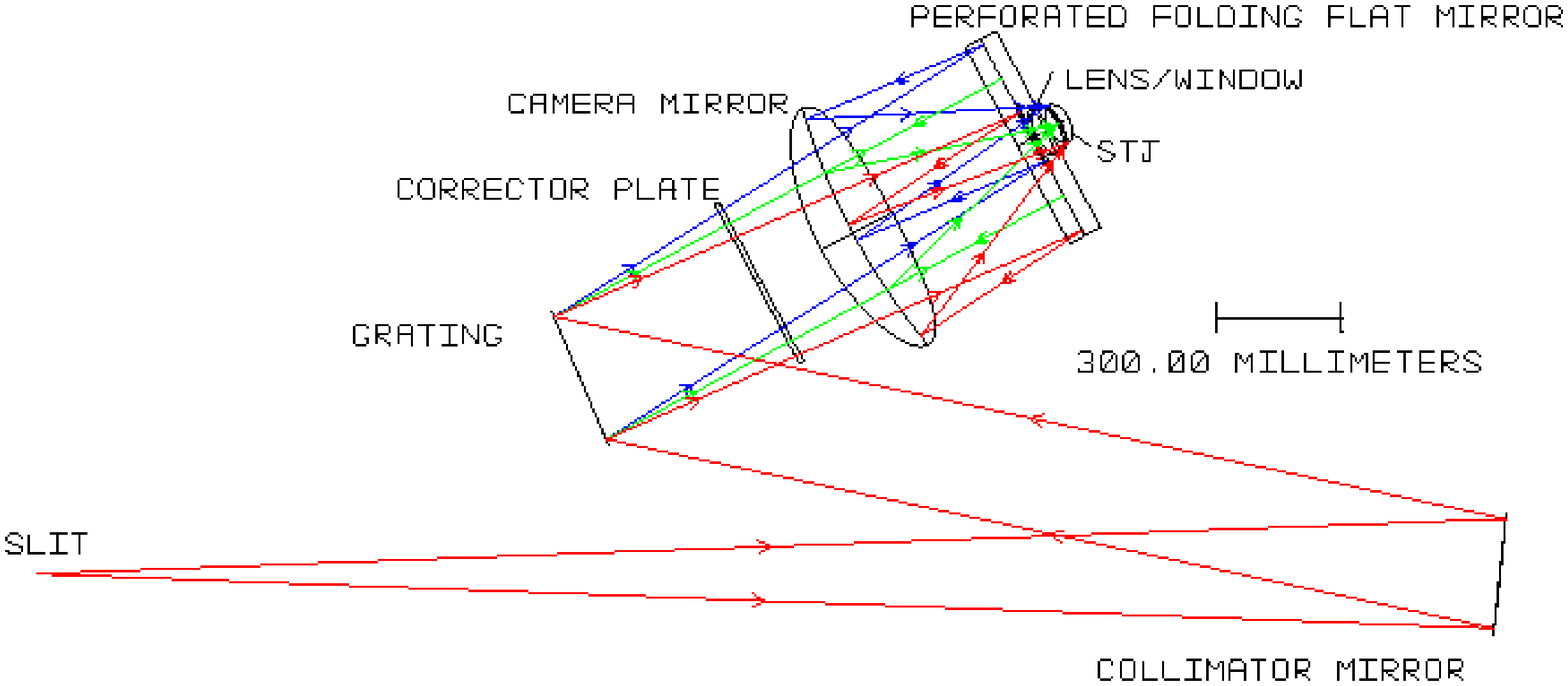}
\\[5mm]
\includegraphics[scale=0.33]{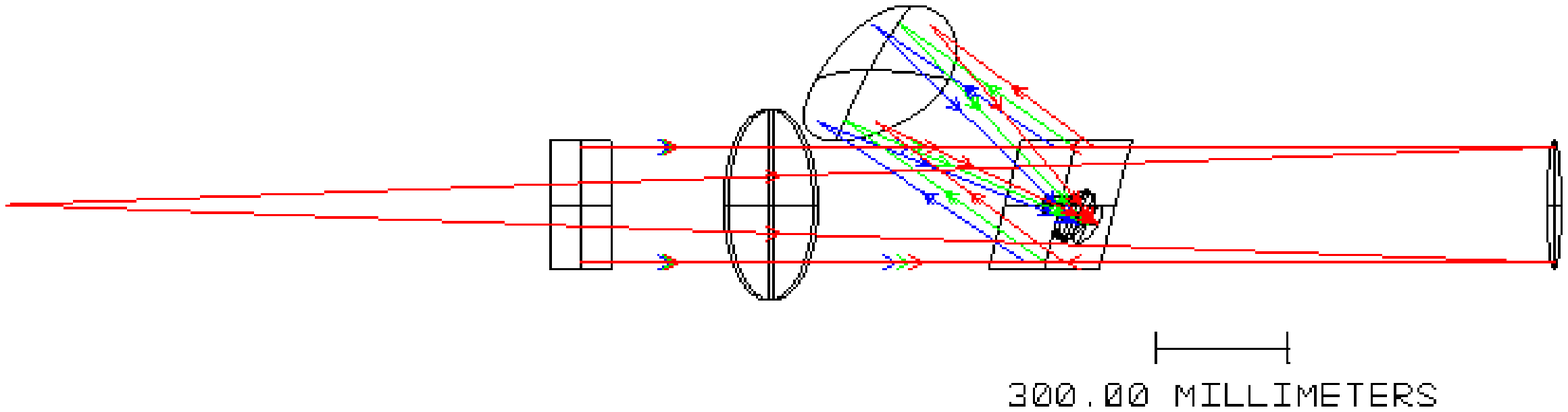}
\end{center}
\caption{Preliminary optical layout. The views are at right angles to
each other. Rays are shown for 1477, 1750 and 2023 nm, the centre and extreme
wavelengths for order 1.}
\label{fig:SES_optics}
\end{figure}

Using the principles in Bingham (1979), we have made a preliminary optical
layout (Figure~\ref{fig:SES_optics}). This is based on an off-axis collimator
of diameter 261 mm feeding a grating of length 319 mm at a $35^{\circ}$ angle
of incidence. The grating has a nominal 278 grooves/mm, blazed at
$15^{\circ}$. Rays are diffracted at $5^{\circ}$ from the grating into a
Schmidt camera of 655 mm focal length ($f/2.2$), with a perforated flat folding
mirror to provide access to the focus. The window of the cryostat would be
figured to flatten the field, and may need to be achromatic.  Optimisation is
required, particularly regarding the camera back focal distance (currently
$\sim 40$ mm) to provide sufficient access to the array within the cryostat,
and as regards thermal aspects (see section~\ref{sec:infrared_issues}).

This preliminary design is relatively straightforward and compact, with the
emphasis on being conservative. It may be possible to increase the
Nyquist-sampled slit width to more closely match the 0.8 arcsec median seeing
typical at most front-rank observatories. This would increase the grating and
optics, in particular the camera, and in order to match the fixed detector
array size a faster camera would be required, which will be more difficult. The
grating size increase can be limited by immersing it, as in the original HROS
concept for Gemini S (D'Arrigo et al 2000).

Other concepts are also possible.  An all-reflecting camera based on a
three-mirror anastigmat (TMA) design may be superior. Off-axis aspheric mirror
technology has made significant strides in the past few years, and such designs
are now routinely being considered. 

Due to atmospheric dispersion, the light from celestial sources is split into
spectra with the blue end pointing towards the zenith. The length of the
spectrum is proportional to the tangent of the zenith distance. In a
traditional echelle spectrograph at Nasmyth or coude, a beam rotator is usually
used to align the atmospheric dispersion along the entrance slit. The
dispersion then adds to, or subtracts from, the cross dispersion. No light is
therefore lost at the aperture, although the echelle orders can `drift'
slightly on the detector during the process of an observation.

While losses will be mitigated by using STJ pixels that are rectangular, it
will be necessary to consider the effect of atmospheric dispersion.  A possible
solution uses two counter-rotating fused-silica prisms (`Risley prisms') ahead
of focus, perhaps with an identical but inverted set near focus. The first
prism-pair compensates for the atmospheric dispersion and provides a
white-light image in the vicinity of the spectrograph entrance
aperture. However, if the aperture were located at this point, the pupil would
be displaced in the vicinity of the spectrograph collimator, and it would also
be dispersed. If this is not acceptable, a second (inverted) pair of prisms is
included, which compensates for both the dispersion and displacement of the
pupil in the spectrograph. The overall beam-offset created by the two
prism-pairs, which varies with zenith distance, is compensated by a telescope
offset.

An atmospheric dispersion compensator of any design in the converging beam
from the telescope would require an aberration-study, and may require curved
surfaces on the prisms. 

\subsection{Infrared background}
\label{sec:infrared_issues}

Tantalum STJs are sensitive to infrared photons at wavelengths beyond the K
band. The sensitivity at wavelengths above $1\mu$m is shown in
figure~8 of Peacock et al (1998) (curve labelled
$E_T$). It decreases to a few percent at $5\mu$m, and then drops precipitously,
before recovering at wavelengths longer than $100\mu$m. 


Infrared photons beyond the range of interest have negative consequences as
follows:
\begin{enumerate}
\item the photons generate only small amounts of charge, so if detected in
isolation, they contribute to the electronic noise peak at low pulse
heights and may advance the tail of this noise into the first order peak;
\item because the energy  of these photons is small, for a given
flux the photon number is high, and the probability of one of these arriving
within the time constant of the pulse-counting electronics is high: this adds a
small amount of charge which broadens the spectral peak and reduces the
resolution;
\item the large photon flux can exceed the maximum count rate of the
pulse-counting electronics, leading to saturation of the device, and
\item the infrared photon flux can induce local heating of the detector
substrate. 
\end{enumerate}
These photons will have been emitted by the source under study, as well as the
thermal environment of the instrument and cryostat, particularly the warm
cryostat window. In S-Cam, stringent measures were taken to eliminate the
infrared flux (Rando et al 2000a), firstly to baffle the field of view seen by
the detector, and then to minimise the flux from the cryostat window and
baffles by using two filters of successive coldness of KG2 glass.  This
arrangement depresses the throughput through the optical by $\sim 40$\%, with a
turnover at $\sim700$ nm, depressing the 1000 nm flux by a factor of
$10^5$. Nevertheless, the remaining infrared flux is still the major
contributor in reducing the spectral resolving power of the STJs from $\sim 20$
to $\sim 12$.

A spectrograph has the advantage over a camera that any flux seen by the
spectrograph at wavelengths longer than the red extreme of order 1 will be
directed to the side of the detector in the direction of the zero order. This
means that the infrared source flux will be less of a problem. However, the
thermal loading from the cryostat window will remain, and, indeed, be more
significant because of the large window size. In addition, because our concept
has an infrared capability, the cutoff of any infrared filter will need to be
at wavelengths longer than the red extreme of order 1. 

In order to keep the thermal background within acceptable limits it will be 
necessary to incorporate the slit and optics grating within a cold environment.
This will allow rejection of all
source {\it and} window-induced flux. It adds to the cost of the instrument 
and influences the cryogenic design significantly, depending strongly on 
whether $\lambda_c(1)$ is in the H band or in the K band.

One technique for unwanted infrared flux rejection is to place the detector at
the focus of a spherical or toroidal mirror, except for an aperture to
accommodate the incoming beam. In this case the detector experiences a radiated
thermal environment approximately appropriate to its surface temperature. This
also provides an opportunity for improving the infrared sensitivity: it may be
possible to refocus reflected infrared source photons back onto the detector,
permitting a second chance for detection. This will require a tilted focal
plane, which is not implemented in the initial optical design in
section~\ref{sec:optical_design}.

\subsection{Calibrations}

The calibration requirements are relatively standard, involving detector
flat-fielding and wavelength calibration. The infrared night sky lines may be
sufficient to provide a continuous monitoring of the wavelength scale for all
wavelengths through the superimposed spectral orders on the detector, but
external lamps would still be required. A calibration unit would therefore be
incorporated, providing an appropriate selection of arc and continuum lamps for
the optical and infrared. The flat-fielding will be wavelength dependent, so
filters would be required to isolate a single order for the flat-field
calibrations. This would also have a blocked position for checks on the (low
levels of) dark noise. The filter unit will be in the main beam so that it can
also be used for astronomical observations, for example should very bright
sources be observed in a single order with very high time resolution.

\subsection{Cryogenics and Magnetic Field Bias}

Tantalum STJs operate at a temperature of 0.3 K. This is just within the reach
of a pumped He3 cryogenic system, but it will probably be better from the point
of view of operating costs to use an unpumped He4 system, with a final sorption
refrigerator stage, as in S-Cam (Verveer et al 1999).  Although the STJ array
will not, in itself, generate significant heat, the parasitic heat injection
through the large harness would indeed be significant. Nevertheless,
considerable experience has been gained in the past decade in such cryogenic
instrumentation on ground-based telescopes, for example the SCUBA submillimeter
bolometer array on the James Clerk Maxwell Telescope on Mauna Kea (0.075K), and
S-Cam2.


\subsubsection{Cryosystem}

Parasitic heat loads on the detector and its cold stage result from radiation
from the surroundings, through the support structure for the cold stage and
through the electrical cryoharness. The first of these imposes a requirement
for nested thermal shrouds. The second requires careful selection of material
and strut design. Materials such as kevlar strings can be considered, but it is
likely that stainless steel struts are more satisfactory mechanically, without
too significant a thermal disadvantage.

The mounting of the STJ array to a support structure would require non-magnetic
materials with low thermal expansion, in order to match the characteristics of
the STJ substrate. Outgassing in this coldest of environments close to
the STJs must also be minimised, as the contaminants will be trapped
preferentially on the coldest part of the system, which includes the detector
array. The material of choice will probably be a ceramic material. Care will
need to be exercised to interface this structure to the cryocooler coldfinger
material.

Significant experience has been gained in the operation of the S-Cam2
cryocooler, and improvements identified, for example in the inner surface
emissivity optimisation, and a reduction in support strut cross-section. This
would be valuable in this spectrograph concept, with the main differences being
a larger focal plane assembly and magnetic bias system, a larger cryostat
window and infrared blocking elements, different constraints on back focal
distance from the optical design and a very much larger parasitic heat loading
through the cryoharness. A particular consideration will be the cooling power
of the He3 sorption refrigerator.

The other general consideration for the cryosystem is the cooling of the optics, 
alluded to in section~\ref{sec:infrared_issues}. This will entail either the 
incorporation of the He4 cryostat within an LN$_2$ or other cryostat, or the 
interfacing of the two cryostats.

\subsubsection{Cryoharness}

For the HSTJ studies we investigated different methods of providing a low
thermal conductivity harness through industrial studies (see Griffiths et al
1997). These included ribbon cables, fine wires, superconducting leads on a
ceramic substrate and thin films on a kapton substrate. It was found that fine
wires made of Manganin or stainless steel, or Niobium tracks laid on kapton
provided acceptable solutions in terms of electrical conductivity and thermal
loading. The surface area, ease of handling and routing, ease of manufacture
and ease of making connections were also important considerations, and these
favoured the Niobium/kapton solution, which also allowed greater uniformity of
the electrical characteristics of the harness (particularly the capacitance). 
There are a small number of common return lines of lower resistance.

The thermal calculations included the conduction along the tracks and through
the kapton, and included radiative loading and losses. In general the
conduction along the tracks exceeds that through the kapton. The effect of
radiation impinging on the cable at temperatures above 20K is important,
requiring the harness to be covered with a low emissivity surface such as a
Gold coating.

The cryoharness must be temperature-clamped on the 4K He4 stage in order to
limit the requirements on the sorption coolers. In practise, for the
spectrograph, the more critical section is the short segment from this stage to
the STJ arrays.  This would be bump-bonded to make the connection to the STJ
detector contacts.

\subsubsection{Magnetic biasing}

A magnet subsystem is required to produce a constant, uniform bias field in
the presence of a possibly magnetically noisy environment during the operation
of the STJ detectors. It must also be possible to vary the
magnetic field during the cooler recycling. This task must be carried out
within the restricted accommodation available in the vicinity of the detectors,
and operate in conjunction with the thermal infrared filtering and baffling in
this locality.

The design of the magnet subsystem depends on the uniformity and stability of
the magnetic field required by the STJ detectors. It also depends on the
magnetic environment of the STJs; in particular, the presence of motor-driven
mechanisms and perhaps compressors in the vicinity of the instrument may impose
a magnetically noisy environment with rapid transients. This would drive the
magnetic field controller time constants to be shorter.

A design adapted from the pre-proposal studies of the magnet subsystem for HSTJ
(Griffiths et al 1997) should be appropriate. This consisted of a screened
Helmholtz coil assembly and a set of magnet power and control electronics. A
scaled-down version of this is used in S-Cam.

The Helmholtz coils are in a $\mu$-metal screening box attached to the open end
of the cryostat. They operate at superconducting temperature to provide
magnetic field bias and trim.  Magnetic field sensors are used to monitor
the stability and uniformity of the magnetic field.

Control electronics would provide interfaces to the magnetic field sensors in
the vicinity of the STJ detectors, control signals for the coil power drive
electronics, an interface to the Instrument Computer and stable and controlled
power to the Helmholtz coils. 

Particular care will need to be taken with contamination. The STJ detectors
would be the coldest elements in the cryostat so contaminants would be
deposited preferentially on them. This would require a careful selection of the
materials to be used in the coils and coil assembly and also some assessment of
the likely contamination paths and rates.

\subsection{Electronics}
\label{sec:concepts_electronics}

Each pixel in the STJ arrays requires its own detector chain. The matrix
readout approach investigated by Martin et al. (2000) has some performance
disadvantages and is not appropriate for the linear arrays required
here, so we have retained the approach of providing an independent
electronic channel for each STJ element. A
$10\,000$-pixel array therefore needs large scale integration in order to
access these pixels, with analog electronics of sufficient quality to
minimally degrade the detector response. Such a task has been
accomplished many times before, for example in high energy collider
instrumentation and even in space applications ({\it Swift}-BAT,
Barthelmy 2000). Using ASICs (Application Specific Integrated Circuits), 
each containing $\sim100$ channels, a
sufficiently low number of circuit cards is needed so they can be
placed close to the detector arays.

\begin{figure}
\begin{center}
\includegraphics[scale=0.45,angle=-90]{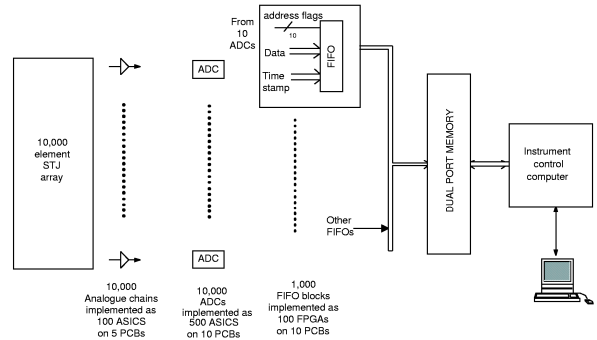}
\end{center}
\caption{The block diagram for the data handling electronics.}
\label{fig:electronics}
\end{figure}

We show in Figure~\ref{fig:electronics} a block diagram for the data-flow
electronics. One
hundred ASICs  each providing around 100 pre-amplifier and analogue 
shaping amplifiers could easily be accommodated on five printed circuit
boards. The Analog-to-Digital Converter (ADC) performance required to convert
the amplified charge packets to digital signals is not demanding due to the
moderate energy resolution of the detectors. An ASIC providing 20
independent 6-bit flash ADCs could easily be developed; 500 of these would be
needed and they could be accommodated on 10 printed circuit boards. These ADCs
would operate asynchronously. A small first-in-first-out buffer (FIFO),
only a few words deep, could be implemented as a Field-Programmable Gate Array
(FPGA) to buffer the output from a group of (say) 10 ADCs and record the event
timestamp. We calculate that the event rate in any group of 10 ADCs would not
exceed the FIFO write rate. Ten such FIFO blocks could easily be contained
within a low cost FPGA. Each FPGA could also buffer the output from its own
FIFOs to provide one output port for onward bussing to a dual port memory which
would be written by the detector system described here. The values written to
this memory would be the 6-bit pixel energy, the 14-bit pixel address and a
32bit timestamp which allows $10\mu$sec to be resolved per 24 hours.

The spectrograph would also require other (relatively standard) instrument
electronics to control and monitor the state of the instrument and cryostat, to
interface with the observatory and telescope systems and to provide a user
interface. Such functions include temperature and magnetic field measurements
and field current control, cryogen level sensing, calibration unit activation
and control, filter wheel operation and slit unit control and viewing. 

\subsection{Structure and mechanical issues}

We envisage the spectrograph to be situated at the Nasmyth focus. Such an
arrangement considerably simplifies the overall structural and mechanical
design. Since the gravity vector is constant, flexure issues are eliminated,
and the cryostat design is made significantly easier.  The spectrograph would
be designed around an optical bench structure, on which are mounted the slit
unit and viewing system, the spectrograph optics and calibration unit and
cryostat. The analog data handling electronics would be
located close to the detector array, with standard interfaces to an external unit containing the majority of instrument control and monitoring
electronics and power supplies. Interfaces for replenishment of the cryogenic
consumables would also need to be considered.

The major mechanical issue would centre on the thermal performance of the
structure, both within the detector He4 cryostat, and, if present, within the
larger cryostat enclosing the spectrograph optics. This is aided by the
almost entirely reflective optical design in our concept. Standard techniques
(for example the use of invar rods) can be used to maintain camera focus
between room and cryogenic temperatures.

The slit unit and slit viewing subsystem are important, but standard items, and
we would not expect these to pose any problems. The same comment applies to the
calibration subsystem. The spectrograph concept would include a filter wheel
located between the slit and collimator, which can be used to introduce
bandpass or neutral density filters or polarisation scramblers if these are
required. These will be useful in commissioning and in calibration: normally
the open position would be used for observations.

\subsection{Control and data handling software}
\label{sec:cdh_software}

The spectrograph would need to operate in the host observatory control and data
handling system.

It is of fundamental importance that the spectrograph provide real-time
feedback not only on the status of the instrument from its various sensors, but
also on the instrument performance directly from the data stream. Outside of an
observation, use of the calibration lamps should allow detector performance,
stability and freedom from electrically-induced noise and also optical
throughtput to be assessed immediately. 

STJ detectors produce a photon event stream characterised by time, position and
energy. As such, their data outputs are more akin to those familiar with data
from X-ray detectors, so that techniques and tools developed in that field
could be the most readily adapted for this concept.  The availability of energy
information makes it simpler to retain the event list format until a final
binning in spectral, temporal or spatial coordinates. A data reduction sequence
could proceed approximately as described in Perryman et al (2001). The data
volume could be large, up to 200 Mbyte/sec, but typically it would be a small
fraction of this.

\section{Performance predictions}

We have developed an instrument simulator to make predictions of the
performance of the spectrograph. 

As inputs we use the standard data for Paranal available from
the ESO ETC (exposure calculator) website: telescope UT1 mirror reflectance, sky
background brightness, extinction, seeing, infrared absorption. The OH sky line
atlas is from Rousselot et al (2000), also used in ETC: this extends down only
to 624 nm, so misses sky lines at shorter wavelengths. The star spectra are
from the atlas of Pickles (1998): these are at $0.5$ nm resolution and extend
over the $320-2500$ nm band of interest.

The simulator interpolates between, or sums over grids of known transmission,
reflectance, emissivity as a function of wavelength, depending on whether the
input grid is more finely or more coarsely sampled that that required for the
prediction. The throughput is calculated per order, then summed to obtain the
overall throughput and S/N ratios, as well as the total count rate on the
array, which is important for ascertaining bright limits.

We limit here the reporting of our exposure estimates to that of the optical+H
band configuration, but the optical+J+K configuration can just as easily be
calculated.

\subsection{Throughput}
\label{sec:throughput}

\begin{figure}
\begin{center}
\includegraphics[scale=0.37]{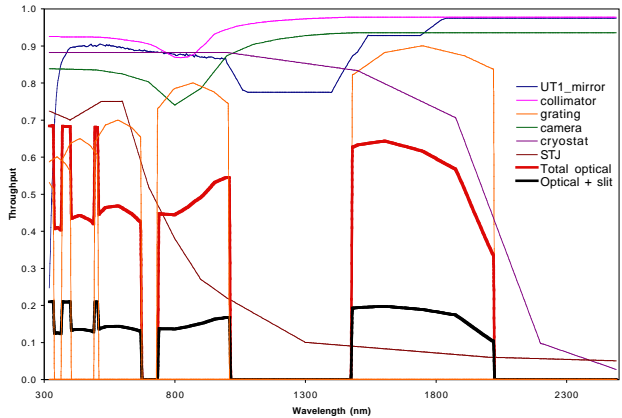}  
\end{center}  
\caption{The throughput of the individual elements of the spectrograph. The
heavy red/grey curve shows the total optical efficiency 
excluding the slit, and the
heavy black curve includes the slit (0.5 arcsec in 0.8 arcsec seeing, array
width 1 arcsec).}
\label{fig:item_efficiency}                        
\end{figure}

We provide details of the throughput of each element of the spectrograph below
the slit for information in Figure~\ref{fig:item_efficiency}. The UT1 mirror
reflectivity is for a single reflection -- 3 reflections are used for
Nasmyth. The throughputs assume the optical design in
section~\ref{sec:optical_design}, and use polarisation-averaged grating
efficiencies for appropriate gratings in gratings catalogs. Also shown is the
total optical efficiency excluding and including the slit.

\begin{figure}
\begin{center}
\includegraphics[bb=120 110 610 470, scale=0.4,angle=0]{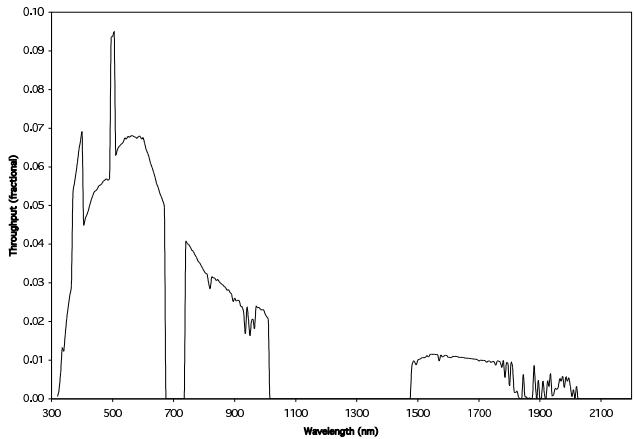} 
\end{center}  
\caption{The overall throughput of the spectrograph including atmospheric 
transmission. Peaks occur where there is order overlap on the array.}
\label{fig:relative_efficiency}                        
\end{figure}

Figure~\ref{fig:relative_efficiency} shows the overall throughput. 
The seeing profile is assumed to be
Gaussian. The calculations use a zenith angle of $0^{\circ}$, seeing of 0.8 arc
sec and a slit width of 0.5 arc sec in the spectral
direction and 1 arcsec in the spatial direction.  The slit is the major source
of losses in the spectrograph: the slit transmission is only $0.3$ for the
above parameters. This is one area where significant improvements may be
possible with optimisations of the optical design (see
section~\ref{sec:optical_design}).

\subsection{Representative Spectra}

\begin{figure}
\begin{center}
\includegraphics[scale=0.37,angle=-90]{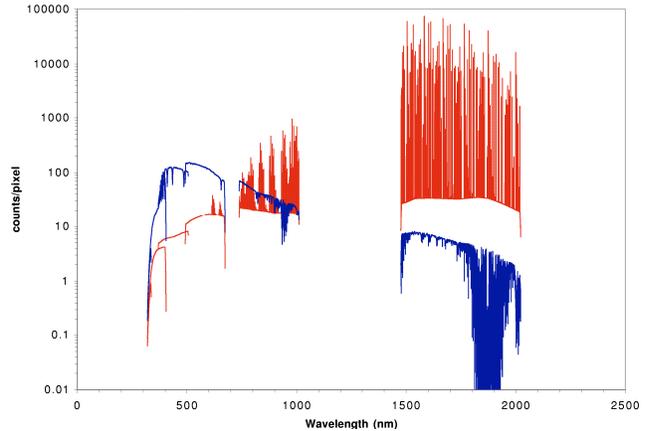}\\[5mm] 
\includegraphics[scale=0.37,angle=-90]{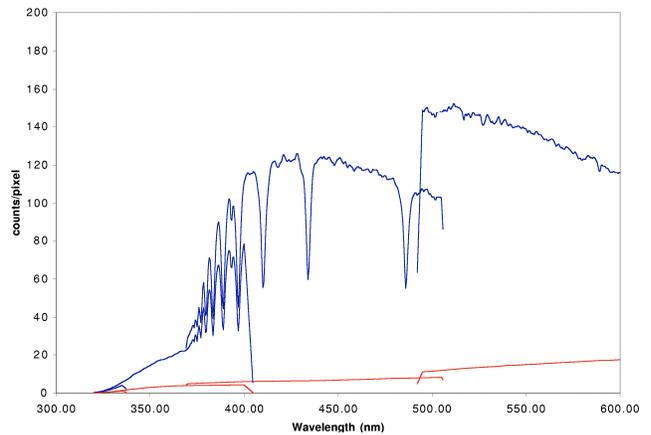} 
\end{center}  
\caption{The total counts spectrum in 1000 sec for a V=20 A0V star 
(blue/black) and the sky background (red/grey), 
assuming a zenith angle of $30^{\circ}$, seeing of
0.8 arcsec and a slit of 0.5 arcsec. The top plot is on a log scale and shows
the full wavelength range, while the lower plot on a linear scale shows an
expanded region in the blue.}
\label{fig:spectrum_A0V}                        
\end{figure}

\begin{figure}
\begin{center}
\includegraphics[scale=0.37,angle=-90]{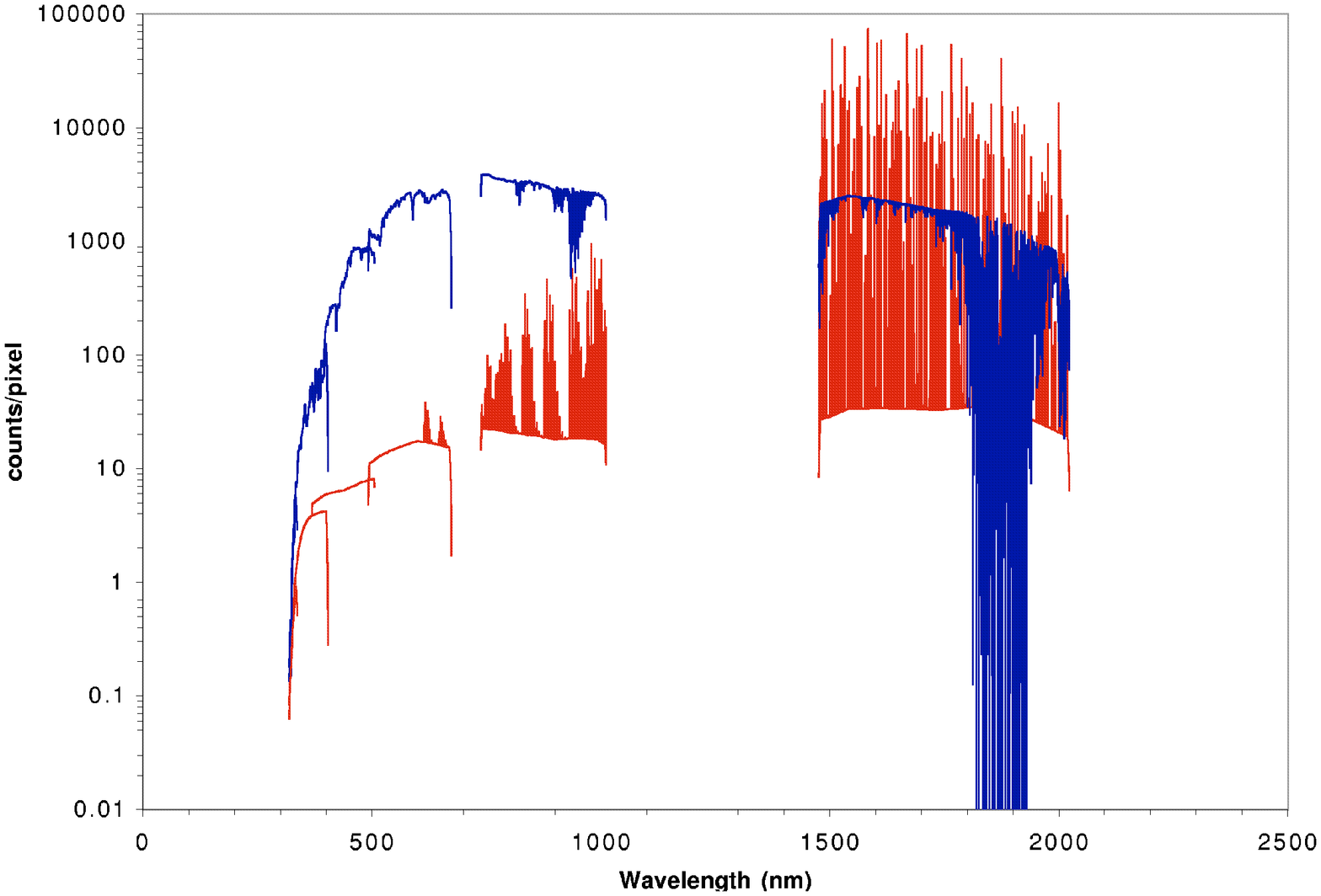}\\[5mm]  
\includegraphics[scale=0.37,angle=-90]{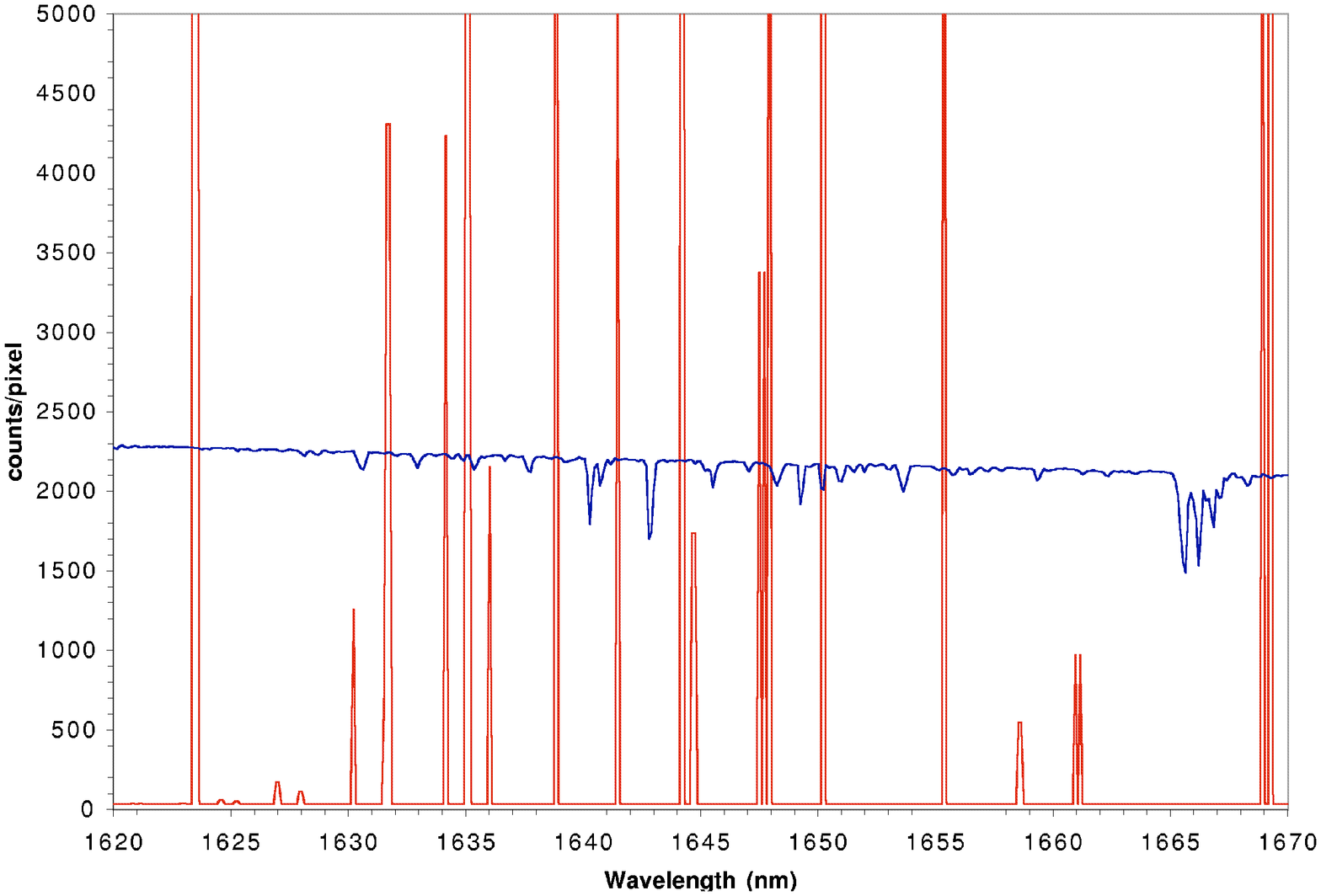}  
\end{center}  
\caption{The total counts spectrum in 1000 sec for a V=17 M0V star 
(blue/black) and the sky background (red/grey), 
assuming a zenith angle of $30^{\circ}$, seeing of
0.8 arcsec and a slit of 0.5 arcsec. The top spectrum of the full wavelength
range is on a log scale, while the lower spectrum, of a short region in the H
band, is on a linear scale.}
\label{fig:spectrum_M0V}
\end{figure}

We have calculated star and sky spectra using the throughputs in
section~\ref{sec:throughput} above. We show in
Figure~\ref{fig:spectrum_A0V} the resulting total count spectrum in 1000
seconds for a V=20
A0V star, and in Figure~\ref{fig:spectrum_M0V} that for a V=17 M0V star (both
from the Pickles atlas). These spectra are noiseless: S/N ratio calculations
are shown later.

For these calculations we assume standard zero points in the literature
(Johnson 1966, Bessel 1979) as used in the ESO ETC, and the collecting area
appropriate for the VLT as given on the the ESO ETC website. The calculations
have been cross-checked against the ESO instrumentation predictions, and found to
be consistent within $20$\%.

The A0V star blue continuum dominates the background even for $V=20$, while in
the red, the night sky lines are a significant component. These sky lines are
seen more clearly in the infrared in the expanded lower plot in
Figure~\ref{fig:spectrum_M0V}, where they are well resolved.

\subsection{Signal to Noise Ratios}
\label{sec:sn-ratios}

\begin{figure}
\begin{center}
\includegraphics[scale=0.37,angle=-90]{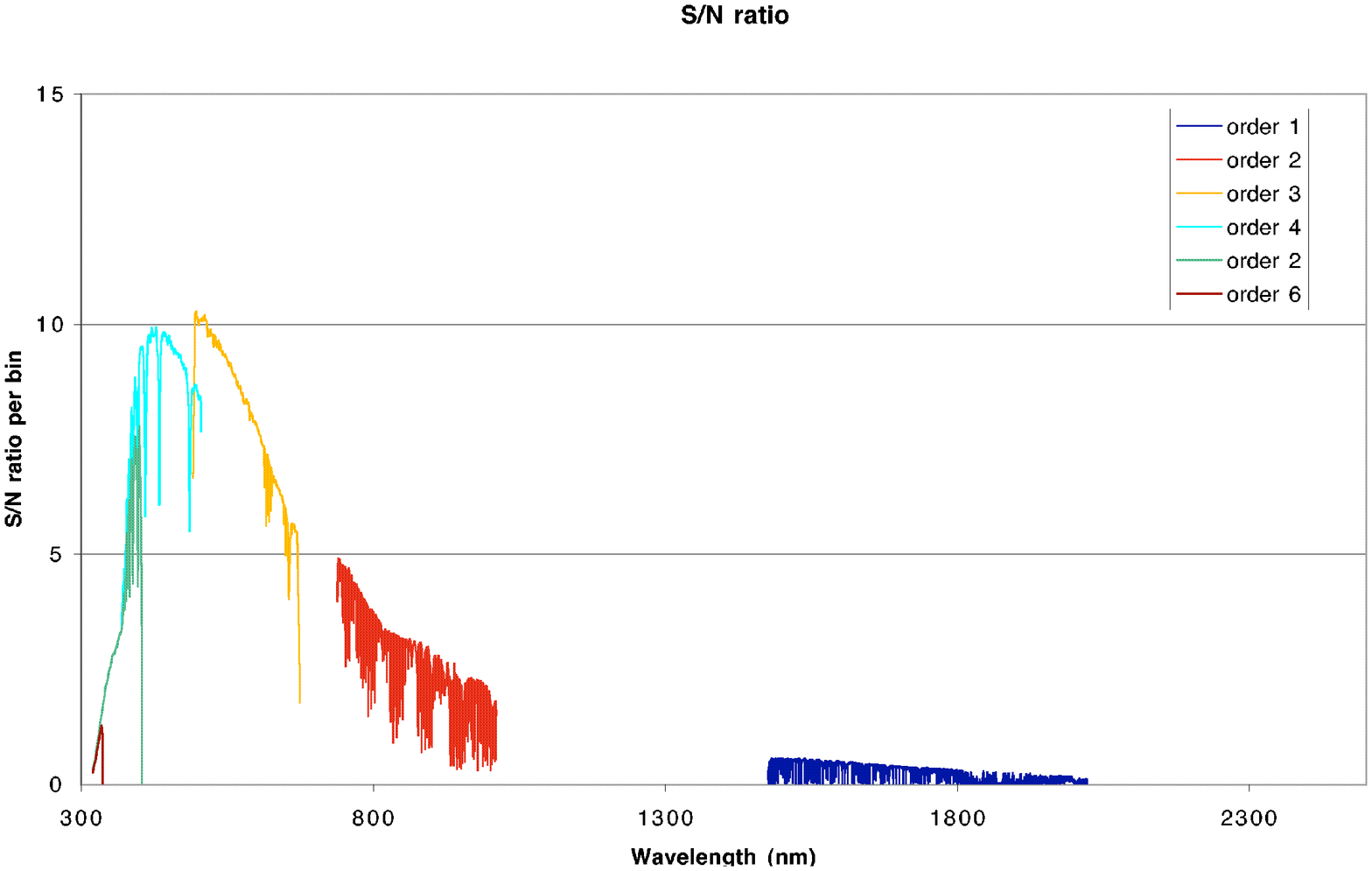}
\includegraphics[scale=0.37,angle=-90]{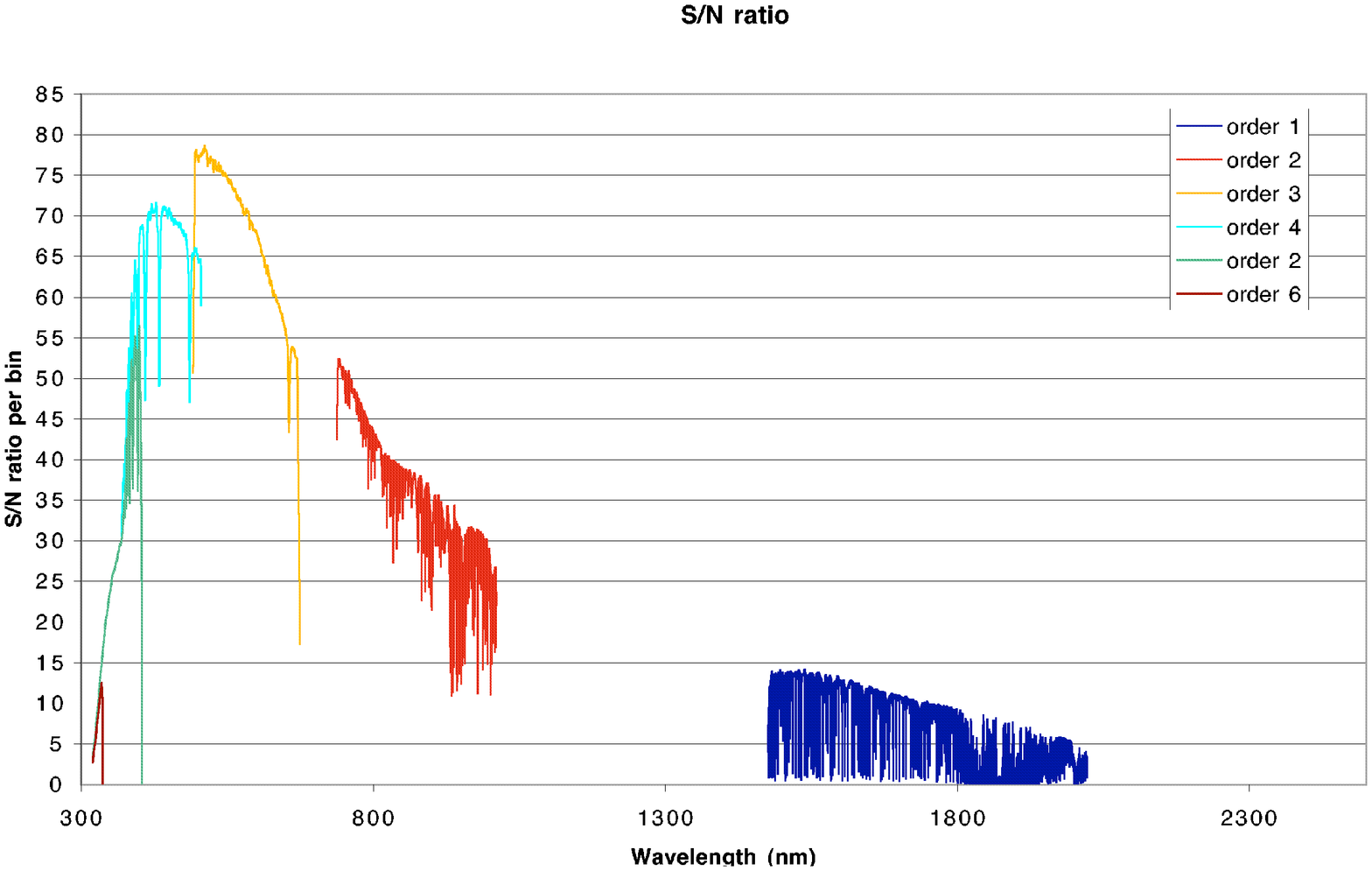}  
\end{center}  
\caption{The signal-to-noise ratios for a $V=22$ (top) and $V=18$
(bottom) A0V star in 3600 sec. The seeing is 0.5 arcsec, the slit is 0.5 arcsec
and the array width 1 arcsec. The zenith angle is $30^{\circ}$. The different
colours/greys indicate different grating orders.}
\label{fig:sn_ratios}
\end{figure}

The S/N ratio calculations use the formulae in the ESO-ETC explanatory notes,
with the simplification that there is no readout noise associated with the
STJs. We show in Figure~\ref{fig:sn_ratios} that the spectrograph provides a
S/N ratio reaching $\sim 10$ per pixel in 3600 sec for an A0V star of $V=22$,
and it reaches this ratio in the H band for $V=H=18$ in the same time (assuming
seeing of 0.5 arcsec). These figures provide a realistic indication as to the
limiting performance at the faint extreme.

Increasing the slit by a factor of 2 increases the throughput through the slit
significantly for the typical seeing conditions at Paranal, at the cost of
reducing the spectral resolving power to 4000.

\subsection{Dynamic range}

\begin{figure}
\begin{center}
\includegraphics[scale=0.37]{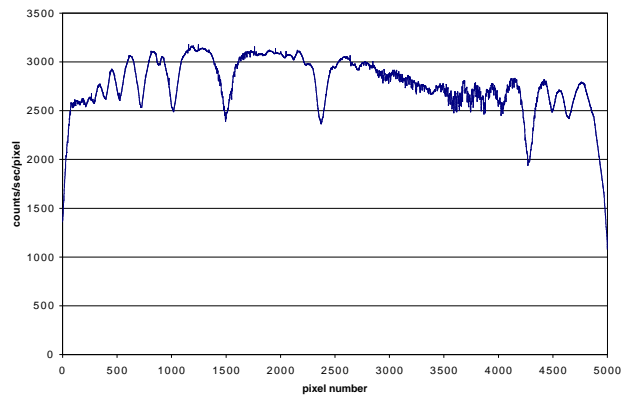}
\end{center}  
\caption{The countrate on the detector array (counts per second per pixel) for
a $V=10$ A0V star. The seeing is 0.8 arcsec with a slit of 0.5 arcsec and an
array width of 1 arcsec; the zenith angle is $30^{\circ}$.}
\label{fig:countrate}
\end{figure}

An important aspect of any photon-counting detector system such as STJs is the
limit set on the dynamic range by the maximum count rate. This is easier to
achieve in a spectroscopic than in an imaging application. We show in
Figure~\ref{fig:countrate} the overall countrate ({\it i.e.} summed over order
number) on the array, as a function of pixel number, for a $V=10$ A0V star. The
countrates reach $\sim3000$ cts/sec/pixel for such a brightness. The current
maximum countrates on S-Cam2 are $\sim5000$ cts/sec/pixel, set by the analog
processing chains. Even if we assume no increase in this capability, this
indicates that the spectrograph will have a high bright limit, sufficient for
good flat-fielding calibrations (already demonstrated in the case of S-Cam2)
and access to celestial spectroscopic standards.

\section{Conclusions}

We have outlined a spectrograph concept which uses the intrinsic wavelength
resolution and extended wavelength response of STJ detectors as the basis for a
high-throughput optical-infrared spectrograph with high time resolution. The
intrinsic wavelength resolution of the detectors is used to perform the order
sorting. Such an STJ-based concept promises an elegantly simple medium
resolution spectrograph. We have calculated that it should be a significant
improvement on existing front-rank instrumentation, opening new regions of
parameter space.

The concept utilises technology which is currently available, or can be
considered to be feasible with only a small level of development.
Nevertheless, it is a step up from existing STJ-based instrumentation, in terms
of array size and wavelength range.  Although we have argued that the
technological increment in moving from existing STJ arrays to those suggested
for this concept is made significantly easier by the fact that the arrays in
this concept are linear, and so provide for straightforward electrical
connectivity, the $2\times2500$ pixel arrays 
would still be a critical development
area. The main challenges are in the area of yield and pixel uniformity (in
terms of electrical characteristics). An additional issue is the buttability.
Another area closely connected with the detector performance is the subsystem
providing the magnetic biasing on the STJ array. The main challenge here is the
uniformity of magnetic field bias, given the physically large arrays. At the
same time the cryostat would provide strong constraints in terms of available
accommodation for the magnets, the siting of infrared baffles and filters, and
the thermal design requirements of the detector support structure.

We have identified further areas where development would be required in the
cryocooler and cryoharness. The issue in the former is the heat pumping
capacity of the sorption cooler(s) to provide the 0.3K temperature drop from
the He4 cryostat. For the latter, the large number of electrical connections to
the detector array mean that the parasitic heat loading through these must be
minimised. There is also the question of the physical routing of several such
cryoharnesses into a relatively small space.

The operation of a $10\,000$-pixel STJ array is 
not feasible without large-scale
integration of the front-end analog electronics. This may require dedicated
ASICs of particular design suitable for STJs. An ameliorating factor is that
the ASICs consist of a large number of relatively simple and independent units.

The infrared baffling and thermal rejection would need careful consideration at
the general instrument level (in terms of rejection of telescope thermal
background) and for the detector environment, in order to minimise the STJ
wavelength resolution degradation as discussed earlier. It has implications for
the overall instrument thermal design, infrared responsivity, on whether the
$\lambda_c(1)$ in the K band option can be exercised, and ultimately on costs.

On the other hand, we have identified the potential for increased infrared
performance by the use of coatings to reduce the reflectivity of Tantalum
beyond 800 nm, and the appropriate design of the thermal infrared rejection
optics around the detector array. A further performance enhancement may be
available through the provision of a degree of spatial resolution along the
slit by the use of DROIDs.

\section{Acknowledgements}

Support for the STJ technological development at ESA-ESTEC is provided by Peter
Verhoeve, Nicola Rando, Didier Martin, Jacques Verveer and Axel van
Dordrecht. Keith Horne is supported by a PPARC Senior Fellowship.


\end{document}